\documentclass{llncs}

\usepackage{wrapfig}

\usepackage{amsmath,latexsym,url}
\usepackage{xspace}
\usepackage{stmaryrd}
\usepackage{tabularx}

\def\lambdot{\rule{0.6mm}{0.6mm}\hspace{0.4ex}} 
\def\all#1{\forall #1\lambdot}
\def\exi#1{\exists #1\lambdot}
\def\lam#1{\lambda #1\lambdot}

\def\modal#1{\boldsymbol{#1}}
\def\mfalse{\modal\bot}
\def\mtrue{\modal\top}
\def\mnot{\modal\neg\,}
\def\mor{\,\modal\vee\,}
\def\mand{\,\modal\wedge\,}
\def\mimpl{\,\modal\supset\,}
\def\miff{\,\modal\Leftrightarrow\,}
\def\mball#1{\modal\Box_{#1}\,}
\def\mdexi#1{\modal\Diamond_{#1}\,}
\def\mall#1{\modal{\forall}{#1}\lambdot\,}
\def\mexi#1{\modal{\exists}{#1}\lambdot\,}

\def\iplemph#1{\dot{#1}}
\def\iplnot{\iplemph\neg\,}
\def\iplor{\,\iplemph\vee\,}
\def\ipland{\,\iplemph\wedge\,}
\def\iplimpl{\,\iplemph\supset\,}

\def\LEO2{{\sc Leo-II}}

\newcommand{\database}[1]{
\rput(0,1.5){\rnode{h1}{\psellipse(1,0)(1,0.3)}}  
\rput(0,0){\psline{-}(0,1.5)}
\rput(2,0){\psline{-}(0,1.5)}
\rput(0,0){\rnode{h2}{\psline[linearc=0.7]{-}(0,0)(0.2,-0.12)(0.6,-0.23)(1,-0.26)(1.4,-0.23)(1.8,-0.12)(2,0)}}
}

\def\ambnormform#1{{#1}\hspace*{-1.1ex}\downarrow_{\kern-.2em\scriptscriptstyle *}} 

\def\IPLSTT{\mathcal{IPL^{STT}}}

\def\QML{\mathcal{QML}}

\def\QMLSTT{\mathcal{QML}^{STT}\xspace}

\def\IVSTT{\mathcal{IV}^{STT}}
\def\PVSTT{\mathcal{PV}^{STT}}
\def\SYMSTT{\mathcal{SYM}^{STT}}
\def\SSTT{\mathcal{S}^{STT}}

\def\STT{\mathcal{STT}\xspace}

\def\QKPIm{\mathbf{QK}\pi^-\xspace}
\def\QKPI{\mathbf{QK}\pi\xspace}
\def\QKPIp{\mathbf{QK}\pi^+\xspace}

\def\stt{\mathcal{STT}\xspace}

\def\ipl{\mathcal{IPL}\xspace}

\def\typearrow{\shortrightarrow}

\def\worldtype{\mu}
\def\indtype{\iota}

\begin{document}
\title{Simple Type Theory as Framework for Combining Logics} 
\author{Christoph  Benzm\"uller\thanks{This work has been funded by the German Research Foundation (DFG) under grant BE 2501/6-1.}}
\institute{
Articulate Software, Angwin, CA, U.S. 
}

\maketitle

\begin{abstract} 
  Simple type theory is suited as framework for 
  combining classical and non-classical logics. This claim is based on
  the observation that various prominent logics, including
  (quantified) multimodal logics and intuitionistic logics, can be elegantly
  embedded in simple type theory. Furthermore, simple type
  theory is sufficiently expressive to model combinations of 
  embedded logics and it has a well understood
  semantics. Off-the-shelf reasoning systems for
  simple type theory exist that can be uniformly employed for
  reasoning {\em within} and {\em about} combinations of logics.
\end{abstract}

\section{Introduction} \label{sec1}
\label{Motivation}
Church's simple type theory $\stt$ \cite{Church40}, also known as
classical higher-order logic, is suited as a framework for combining
classical and non-classical logics. This is what this paper illustrates. 

Evidently, $\stt$ has many prominent classical logic fragments,
including propositional and first-order logic, the guarded fragment,
second-order logic, monadic second-order logic, the basic fragment of
$\stt$, etc.  Interestingly, also prominent non-classical logics --
including quantified multi-modal logics and intuitionistic logic --
can be elegantly embedded in $\stt$. It is thus not surprising that
also combinations of such logics can be flexibly modeled within
$\stt$.  Our claim is furthermore supported by the fact that the
semantics of $\stt$ is well understood
\cite{Andrews72b,Andrews72a,BBK04,Henkin50} and that powerful proof
assistants and automated theorem provers for $\stt$ already exist.
The automation of $\stt$ currently experiences a renaissance that has
been fostered by the recent extension of the successful TPTP
infrastructure for first-order logic
\cite{DBLP:journals/jar/Sutcliffe09} to higher-order logic, called
TPTP THF \cite{BRS08,C29}. Exploiting this new infrastructure we will
demonstrate how higher-order automated theorem provers and
model generators can be employed for reasoning {\em within} and {\em
  about} combinations of logics.

In Sect. \ref{qml} we outline our embedding of quantified multimodal
logics in $\stt$.  Further logic embeddings in $\stt$ are discussed in
Sect. \ref{sec:intui}; our examples comprise intuitionistic logic, access
control logics and the region connection calculus.  In
Sect. \ref{sec:about} we illustrate how the reasoning \textit{about}
logics and their combinations is facilitated in our approach, and in
Sect. \ref{sec:within} we employ simple examples to demonstrate the
application of our approach for reasoning \textit{within} combined
logics.  The performance results of our experiments with
off-the-shelf, TPTP THF compliant higher-order automated reasoning
systems are presented in Sect. \ref{sec:exp}.

\section{(Normal) Quantified Multimodal Logics in $\stt$} \label{qml}

$\stt$ \cite{Church40} is based on the simply typed
$\lambda$-calculus. The set~$\mathcal{T}$ of simple types is usually
freely generated from a set of basic types $\{o, \indtype\}$ (where
$o$ is the type of Booleans and $\indtype$ is the type of individuals)
using the right-associative function type constructor
$\typearrow$. Instead of $\{o, \indtype\}$ we here consider a set of
base types $\{o, \indtype, \worldtype\}$, providing an additional base
type $\worldtype$ (the type of possible worlds).

The simple type theory language $\stt$ is defined by (where $\alpha$,
$\beta$, $o\in\mathcal{T}$):
\begin{eqnarray*} 
s,t & ::= & p_\alpha  \mid X_\alpha  \mid (\lam{X_\alpha} s_\beta)_{\alpha\typearrow\beta} \mid (s_{\alpha\typearrow\beta}\, t_\alpha)_\beta \mid (\neg_{o\typearrow o}\, s_o)_o \mid \\ & & (s_o \vee_{o\typearrow o \typearrow o} t_o)_o \mid 
(s_\alpha =_{\alpha\typearrow\alpha\typearrow o} t_\alpha)_o
\mid (\Pi_{(\alpha\typearrow o)\typearrow o}\, s_{\alpha\typearrow o})_o 
\end{eqnarray*}
$p_\alpha$ denotes typed constants and $X_\alpha$ typed variables
(distinct from $p_\alpha$).  Complex typed terms are constructed via
abstraction and application.  Our logical connectives of choice are
$\neg_{o\typearrow o}$, $\lor_{o\typearrow o\typearrow o}$,
$=_{\alpha\typearrow\alpha\typearrow o}$ and $\Pi_{(\alpha\typearrow
  o)\typearrow o}$ (for each type $\alpha$).\footnote{This choice is
  not minimal (from
  $=_{\alpha\typearrow\alpha\typearrow o}$ all other logical constants
  can already be defined \cite{Andrews2002a}). It useful though in the context of resolution based theorem
  proving.}  From these connectives, other logical
connectives can be defined in the usual way (e.g., $\wedge$ and
$\Rightarrow$). We often use binder notation $\all{X_\alpha} s$ for
$\Pi_{(\alpha\typearrow o)\typearrow o}(\lam{X_\alpha} s_o)$.  
We assume familiarity with $\alpha$-conversion, $\beta$- and
$\eta$-reduction, and the existence of $\beta$- and $\beta\eta$-normal
forms. Moreover, we obey the usual definitions of free variable
occurrences and substitutions.
 
The semantics of $\stt$ is well understood and thoroughly documented
in the literature \cite{Andrews72b,Andrews72a,BBK04,Henkin50}. The
semantics of choice for our work is Henkin semantics.

Quantified modal logics have been studied by Fitting \cite{Fitting02}
(further related work is available by Blackburn and Marx
\cite{BlackburnMarx02} and Bra\"uner \cite{Brauner05}). In contrast to
Fitting we are here not interested only in \textbf{S5} structures but
in the more general case of \textbf{K} from which more constrained
structures (such as \textbf{S5}) can be easily obtained. First-order
quantification can be constant domain or varying domain.  Below we
only consider the constant domain case: every possible world has the
same domain. Like Fitting, we keep our definitions simple by not
having function or constant symbols.  While Fitting \cite{Fitting02}
studies quantified monomodal logic, we are interested in quantified
multimodal logic. Hence, we introduce multiple $\mball{r}$ operators
for symbols $r$ from an index set $S$. The grammar for our quantified
multimodal logic $\QML$ hence is
\begin{align*} 
  s,t & ::=  P \mid k(X^1,\ldots,X^n) \mid \mnot s \mid s \mor t \mid \all{X} s  \mid \all{P} s  \mid \mball{r} s 
\end{align*}
where $P$ denotes propositional variables, $X,X^i$ denote first-order
(individual) variables, and $k$ denotes predicate symbols of any arity.
Further connectives, quantifiers, and modal operators can be defined
as usual.  We also obey the usual definitions of free variable
occurrences and substitutions.

Fitting introduces three different notions of Kripke semantics for $\QML$:
\textbf{QS5$\pi^-$}, \textbf{QS5$\pi$}, and \textbf{QS5$\pi^+$}.  In our work \cite{R45}
we study related notions $\QKPIm$, $\QKPI$, and $\QKPIp$ for a modal
context \textbf{K}, and we support multiple modalities.

$\stt$ is an expressive logic and it is thus not surprising that
$\QML$ can be elegantly modeled and even automated as a fragment of
$\stt$.  The idea of the encoding, called {$\QMLSTT$}, is
simple. Choose type $\indtype$ to denote the (non-empty) set of
individuals and we reserve a second base type $\worldtype$ to denote
the (non-empty) set of possible worlds.  The type $o$ denotes the set
of truth values.  Certain formulas of type $\worldtype\typearrow o$
then correspond to multimodal logic expressions. The multimodal
connectives $\modal\neg$, $\modal\vee$, and $\modal\Box$, become $\lambda$-terms of
types ${(\worldtype\typearrow o)\typearrow(\worldtype\typearrow o)}$,
${(\worldtype\typearrow o)\typearrow(\worldtype\typearrow o)
  \typearrow (\worldtype\typearrow o)}$, and
${(\worldtype\typearrow\worldtype\typearrow
  o)\typearrow(\worldtype\typearrow o) \typearrow (\worldtype
  \typearrow o)}$ respectively.

Quantification is handled as in $\stt$ by modeling
$\all{X} p$ as $\Pi(\lam{X}.p)$ for a suitably chosen connective
$\Pi$.  Here we are interested in defining two
particular modal $\modal\Pi$-connectives: $\modal\Pi^\indtype$, for
quantification over individual variables, and
$\modal\Pi^{\worldtype\typearrow o}$, for quantification over modal
propositional variables that depend on worlds. They become terms of
type $(\indtype\typearrow(\worldtype \typearrow o))\typearrow(\worldtype \typearrow
o)$ and $((\worldtype\typearrow o)\typearrow(\worldtype \typearrow
o))\typearrow(\worldtype \typearrow o)$ respectively.


The $\QMLSTT$ modal operators $\modal\neg,\modal\vee,\modal\Box,\modal\Pi^\indtype$, and $\modal\Pi^{\worldtype \typearrow o}$ are now simply defined as follows:
\begin{align*}
\mnot_{(\worldtype\typearrow o)\typearrow(\worldtype\typearrow o)} & = \lam{\phi_{\worldtype\typearrow o}}\lam{W_\worldtype}\neg \phi\,W\\
\mor_{(\worldtype\typearrow o)\typearrow(\worldtype\typearrow o)\typearrow(\worldtype\typearrow o)} & = \lam{\phi_{\worldtype\typearrow o}} \lam{\psi_{\worldtype\typearrow o}} \lam{W_\worldtype} \phi\,W \vee \psi\,W\\
\mball{}_{(\worldtype\typearrow\worldtype\typearrow o)\typearrow(\worldtype\typearrow o)\typearrow(\worldtype\typearrow o)} & =  
\lam{R_{\worldtype\typearrow\worldtype\typearrow o}} \lam{\phi_{\worldtype\typearrow o}} \lam{W_{\worldtype}} \all{V_{\worldtype}} \neg R\,W\,V \vee \phi\,V \\
\modal\Pi^\indtype_{(\indtype\typearrow(\worldtype \typearrow o))\typearrow(\worldtype \typearrow o)} & = \lam{\phi_{\indtype\typearrow(\worldtype \typearrow
    o)}} \lam{W_\worldtype} \all{X_\indtype} \phi\,X\,W \\
\modal\Pi^{\worldtype \typearrow o}_{((\worldtype \typearrow o)\typearrow(\worldtype \typearrow o))\typearrow(\worldtype \typearrow o)} & = \lam{\phi_{(\worldtype \typearrow o)\typearrow(\worldtype \typearrow o)}} \lam{W_\worldtype} \all{P_{\worldtype\typearrow o}} \phi\,P\,W
\end{align*}

Note that our encoding actually only employs the second-order fragment of $\stt$ enhanced with lambda-abstraction.

Further operators can be introduced as usual, for example, $\mtrue =
\lam{W_\worldtype} \top, \mfalse = \mnot \mtrue$, $\mand =
\lam{\phi,\psi} \mnot (\mnot \phi \mor \mnot \psi)$, $\mimpl =
\lam{\phi,\psi} \mnot \phi \mor \psi$, $\miff = \lam{\phi,\psi} (\phi
\mimpl \psi) \wedge (\psi \mimpl \phi)$, $\mdexi{} = \lam{R,\phi}
\mnot (\modal\Box\,R\,(\mnot \phi))$, $\modal\Sigma^\indtype =
\lam{\phi} \mnot\modal\Pi^\indtype (\lam{X} \mnot \phi\, X)$,
$\modal\Sigma^{\worldtype \typearrow o} = \lam{\phi}
\mnot\modal\Pi^{\worldtype \typearrow o} (\lam{P} \mnot \phi\, P)$.

For defining $\QMLSTT$-propositions we fix a set $\IVSTT$ of individual variables of type $\indtype$, a set $\PVSTT$
of propositional variables\footnote{Note that the denotation of propositional variables depends on worlds.} of type $\worldtype\typearrow o$, and a set $\SYMSTT$ of $n$-ary (curried) predicate constants of types ${\text{${\underbrace{\indtype\typearrow\ldots \typearrow\indtype}_{n}\typearrow(\worldtype\typearrow o)}$}}$. 
Moreover, we fix a set $\SSTT$ of accessibility relation constants of type $\worldtype\typearrow \worldtype \typearrow o$. $\QMLSTT$-propositions are now defined as the smallest set of $\stt$-terms for  which the following hold:
\begin{itemize}
\item if $P\in\PVSTT$, then $P\in\QMLSTT$
\item if $X^j\in\IVSTT$ ($j=1$, \ldots, $n$) and $k\in\SYMSTT$, then $(k\,X^1 \, \ldots \, X^n) \in \QMLSTT$
\item if $\phi,\psi \in \QMLSTT$, then $\mnot\,\phi\in \QMLSTT$ and 
  $\phi\mor\psi\in \QMLSTT$
\item if $r\in\SSTT$ and $\phi\in\QMLSTT$, then $\modal\Box\,r\,\phi\in\QMLSTT$. 
\item if $X\in\IVSTT$ and $\phi\in\QMLSTT$, then  $\modal\Pi^\indtype(\lam{X} \phi)\in\QMLSTT$
\item if $P\in\PVSTT$ and $\phi\in\QMLSTT$, then  $\modal\Pi^{\worldtype\typearrow o}(\lam{P} \phi)\in\QMLSTT$ 
\end{itemize}
We write
  $\mball{r}\phi$ for $\modal\Box\,r\,\phi$, $\all{X_\indtype}\phi$ for $\modal\Pi^\indtype(\lam{X_\indtype} \phi)$, and  $\all{P_{\worldtype\typearrow o}}\phi$ for $\modal\Pi^{\worldtype\typearrow o}(\lam{P_{\worldtype\typearrow o}} \phi)$.

  Note that the defining equations for our $\QML$ modal operators are
  themselves formulas in $\stt$. Hence, we can express
  $\QML$ formulas in a higher-order reasoner elegantly in the
  usual syntax. For example, $\mball{r} \mexi{P_{\worldtype\typearrow
      o}} P$ is a $\QMLSTT$ proposition; it has type
  ${\worldtype\typearrow o}$.

  Validity of $\QMLSTT$
  propositions is defined in the obvious way: a $\QML$-proposition
  $\phi_{\worldtype\typearrow o}$ is valid if and only if for all
  possible worlds $w_\worldtype$ we have $w\in
  \phi_{\worldtype\typearrow o}$, that is, if and only if
  $\phi_{\worldtype\typearrow o}\,w_\worldtype$ holds. Hence, the notion of 
  validity is modeled via the following equation (alternatively we could define $\text{valid}$ simply as $\Pi_{(\worldtype\typearrow o) \typearrow o}$):
 \[\text{valid} =  \lam{\phi_{\worldtype\typearrow o}} \all{W_{\worldtype}} \phi\,W \]

 Now we can formulate proof problems in $\QMLSTT$, e.g., $\text{valid}
 \, \mball{r} \mexi{P_{\worldtype\typearrow o}} P$.  Using rewriting
 or definition expanding, we can reduce such proof problems to
 corresponding statements containing only the basic connectives
 $\neg$, $\vee$, $=$, $\Pi^\indtype$, and $\Pi^{\worldtype\typearrow
   o}$ of $\stt$. In contrast to the many other approaches no external
 transformation mechanism is required.  For our example formula
 $\text{valid} \, \mball{r} \mexi{P_{\worldtype\typearrow o}} P$
 unfolding and $\beta\eta$-reduction leads to $\all{W_\worldtype}
 \all{Y_\worldtype} \neg r\,W\,Y \vee (\neg
 \all{X_{\worldtype\typearrow o}} \neg (X\, Y))$.  It is easy to check
 that this formula is valid in Henkin semantics: put $X =
 \lam{Y_\worldtype} \mtrue$.

 We have proved soundness and completeness for this embedding
 \cite{R45}, that is, for $s\in\QML$ and the corresponding $s_{\worldtype\typearrow o} \in \QMLSTT \subset \stt$ we have:

\begin{theorem} \label{thm1}
$\models^{\stt}
  (\text{valid}\, s_{\worldtype\typearrow o})$  if and only if $\models^{\QKPI} s$. 
\end{theorem}

This result also illustrates the correspondence between
${\QKPI}$ models and Henkin models; for more details see \cite{R45}.

Obviously, the reduction of our embedding  to first-order multimodal logics (which only allow quantification over individual variables), to propositional quantified multimodal logics (which only allow quantification over propositional variables) and to propositional multimodal logics (no
quantifiers)
is sound and complete. Extending our embedding for hybrid logics is straightforward
\cite{KaSmoJoLLI2009}; note in particular that denomination of individual worlds using constant symbols of type $\worldtype$ is easily possible.


In the remainder we will often omit type information. It is sufficient to remember that
worlds are of type $\worldtype$, multimodal propositions of type $\worldtype \typearrow o$, and accessibility
relations of type  $\worldtype \typearrow \worldtype \typearrow o$. Individuals are of type $\indtype$.


\section{Embeddings of Other Logics in $\stt$} \label{sec:intui} We
have studied several other logic embeddings in $\stt$, some of which
will be mentioned in this section.

\paragraph{Intuitionistic Logics} 
G\"{o}dels interpretation of propositional intuitionistic logic in
propositional modal logic $S4$ \cite{Goedel33} can be combined with
our results from the previous section in order to provide a sound and
complete embedding of propositional intuitionistic logic into $\stt$ \cite{R45}.

G\"odel studies the propositional intuitionistic logic $\ipl$ defined by
$$ s,t ::= p \mid \iplnot s \mid s \iplimpl t \mid s \iplor t \mid p \ipland t $$

He introduces the a mapping from $\ipl$ into propositional modal logic
$S4$ which maps $\iplnot s$ to $\mnot \mball{r} s$, $s \iplimpl t$ to
$\mball{r} s \mimpl \mball{r} t$, $s \iplor t$ to $\mball{r} s \mor
\mball{r} t$, and $s \ipland t$ to $s \mand t$.\footnote{Alternative mappings have been proposed and studied in the literature
which we could employ here equally as well.}  By simply combining
G\"odel's mapping with our mapping from before we obtain the following
embedding of $\ipl$ in $\stt$.

Let $\ipl$ be a propositional intuitionistic logic with atomic
primitives $p^1$, \ldots, $p^m$ $(m \geq 1)$ .
We define the set $\IPLSTT$ of corresponding propositional intuitionistic logic propositions in $\stt$ as follows.
\begin{enumerate}
\item For the atomic $\ipl$ primitives $p^1$, \ldots, $p^m$ we introduce corresponding $\IPLSTT$ predicate constants $p^1_{\worldtype\typearrow o}$, \ldots, $p^m_{\worldtype\typearrow o}$. Moreover, we provide the single accessibility relation constant $r_{\worldtype\typearrow\worldtype\typearrow o}$.
\item Corresponding to G\"odel's mapping we introduce the logical connectives of $\IPLSTT$ as abbreviations for the following $\lambda$-terms (we omit the types here): 
\begin{eqnarray*}
\iplnot & = & \lam{\phi}\lam{W}\neg \all{V} \neg r\,W\,V \vee \phi\,V\\
\iplimpl & = & \lam{\phi} \lam{\psi} \lam{W} \neg (\all{V} \neg r\,W\,V \vee \phi\,V) \vee (\all{V} \neg r\,W\,V \vee \psi\,V)\\
\iplor & = & \lam{\phi} \lam{\psi} \lam{W} (\all{V} \neg r\,W\,V \vee \phi\,V) \vee (\all{V} \neg r\,W\,V \vee \psi\,V)\\
\ipland  & = &  \lam{\phi} \lam{\psi} \lam{W} \neg (\neg \phi\,W \vee \neg \psi\,W)
\end{eqnarray*}

\item We define the set of $\IPLSTT$-propositions as the smallest set of simply typed $\lambda$-terms for which the following hold:
\begin{itemize}
\item $p^1_{\worldtype\typearrow o}$, \ldots, $p^m_{\worldtype\typearrow o}$  define the atomic $\IPLSTT$-propositions.
\item If $\phi$ and $\psi$ are $\IPLSTT$-propositions, then so are $\iplnot\,\phi$, $\phi\iplimpl\psi$, $\phi\iplor\psi$,  and  $\phi\ipland\psi$.
\end{itemize}
\end{enumerate}

The notion of validity we adopt is the same as for $\QMLSTT$. However, since G\"odel connects $\ipl$ with
modal logic $S4$, we transform each proof problem $t\in\ipl$ into a corresponding
proof problem $t'$ in $\stt$ of the following form
$$
t' := ((\text{valid}\; \mall{\phi_{\worldtype\typearrow o}} \mball{r} \phi\mimpl \phi) \wedge
(\text{valid}\,  \mall{\phi_{\worldtype\typearrow o}} \mball{r} \phi \mimpl
\mball{r} \mball{r} \phi)) \Rightarrow (\text{valid}\, t_{\worldtype\typearrow o})
$$
where $t_{\worldtype\typearrow o}$ is the $\IPLSTT$ term for $t$ according
to our definition above. Alternatively we may translate $t$ into 
$
t'' := ((\text{reflexive}\, r) \wedge (\text{transitive}\, r)) \Rightarrow (\text{valid}\, t_{\worldtype\typearrow o})
$.

Combining soundness \cite{Goedel33}  and completeness \cite{mckinsey48} of G\"odel's embedding with 
Theorem \ref{thm1} we obtain the following soundness and completeness result: 
Let $t\in\ipl$ and let $t'\in\stt$ as constructed above. $t$ is valid
in propositional intuitionistic logic if and only if $t'$ is valid in $\stt$.

Example problems in intuitionistic logic have been encoded in THF syntax \cite{BRS08} and added to the TPTP THF library\footnote{TPTP THF problems for various problem categories are available at \url{http://www.cs.miami.edu/~tptp/cgi-bin/SeeTPTP?Category=Problems}; all problem identifiers with an '$\hat{\;}$' in their name refer to higher-order THF problems. The TPTP library meanwhile contains more than 2700 example problems in THF syntax.}
and are accessible under identifiers 
{SYO058$\hat{\;}$4} -- \text{SYO074$\hat{\;}$4}.

\paragraph{Access Control Logics}
Garg and Abadi recently translated several prominent access control
logics into modal logic S4 and proved these translations sound and
complete \cite{GargAbadi08}. We have combined this work with our above
results in order to obtain a sound and complete embedding of these
access control logics in $\stt$ and we have carried out experiments
with the prover LEO-II \cite{C27}. Example problems have been
added to the TPTP THF library and are accessible under identifiers 
SWV425$\hat{\;}x$ -- SWV436$\hat{\;}x$ (for $x\in\{1,\ldots,4\}$).

\paragraph{Logics for Spatial Reasoning} Evidently, the region
connection calculus \cite{DBLP:conf/kr/RandellCC92} is a fragment of
$\stt$: choose a base type $r$ ('region') and a reflexive and
symmetric relation $c$ ('connected') of type $r \typearrow r
\typearrow o$ and define (where $X,Y,$ and $Z$ are variables of type
$r$):
\begin{eqnarray*}
\text{disconnected}: & dc & = \lam{X,Y} \neg (c\; X\; Y) \\
\text{part of}: & p & = \lam{X,Y} \all{Z} ((c\; Z\; X) \Rightarrow (c\; Z\; Y)) \\
\text{identical with}: & eq & = \lam{X,Y} ((p\; X\; Y) \wedge (p\; Y\; X)) \\
\text{overlaps}: & o & = \lam{X,Y} \exi{Z} ((p\; Z\; X) \wedge (p\; Z\; Y)) \\
\text{partially overlaps}: & po & = \lam{X,Y} ((o\; X\; Y) \wedge \neg (p\; X\; Y) \wedge \neg (p\; Y\; X)) \\
\text{externally connected}: & ec & = \lam{X,Y} ((c\; X\; Y) \wedge \neg (o\; X\; Y)) \\
\text{proper part}: & pp & = \lam{X,Y} ((p\; X\; Y) \wedge \neg (p\; Y\; X)) \\
\text{tangential proper part}: & tpp & = \lam{X,Y} ((pp\; X\; Y) \wedge \exi{Z} ((ec\; Z\; X) \wedge (ec\; Z\; Y))) \\
\text{nontang. proper part}: & ntpp & = \lam{X,Y} ((pp\; X\; Y) \wedge \neg \exi{Z} ((ec\; Z\; X) \wedge (ec\; Z\; Y))) 
\end{eqnarray*}
An example problem for the region connection calculus will be discussed below.

\section{Reasoning about Logics and Combinations of
  Logics} \label{sec:about} We illustrate how our approach supports
reasoning about logics and their combinations. First, we focus on modal
logics and their well known relationships between properties of
accessibility relations and corresponding modal axioms (respectively axiom
schemata)  \cite{goldblatt92}. Such meta-theoretic insights can be elegantly encoded (and,
as we will later see, automatically proved) in our approach.  First we
encode various accessibility relation properties in $\stt$:
\begin{eqnarray}
\text{reflexive} & = & \lam{R}\all{S} R\,S\,S \label{15} \\
\text{symmetric} & = & \lam{R}\all{S,T} ((R\,S\,T) \Rightarrow (R\,T\,S)) \label{16} \\
\text{serial} & = & \lam{R}\all{S} \exi{T} (R\,S\,T) \label{17} \\
\text{transitive} & = & \lam{R}\all{S,T,U} ((R\,S\,T) \wedge (R\,T\,U) \Rightarrow (R\,S\,U))\label{18} \\
\text{euclidean} & = & \lam{R}\all{S,T,U} ((R\,S\,T) \wedge (R\,S\,U) \Rightarrow (R\,T\,U))\label{19} \\
\text{partially\_functional} & = & \lam{R}\all{S,T,U} ((R\,S\,T) \wedge (R\,S\,U) \Rightarrow T = U)\label{20} \\
\text{functional} & = & \lam{R}\all{S} \exi{T} ((R\,S\,T) \wedge \all{U} ((R\,S\,U) \Rightarrow T = U)) \label{21} \\
\text{weakly\_dense} & = & \lam{R}\all{S,T} ((R\,S\,T) \Rightarrow \exi{U} ((R\,S\,U) \wedge (R\,U\,T))) \label{22} \\
\text{weakly\_connected} & = & \lam{R}\all{S,T,U} (((R\,S\,T) \wedge (R\,S\,U)) \Rightarrow   \nonumber \\
 & & \,\,\, ((R\,T\,U) \vee T = U \vee (R\,U\,T))) \label{23} \\ 
\text{weakly\_directed} & = &  \lam{R}\all{S,T,U} (((R\,S\,T) \wedge (R\,S\,U)) \Rightarrow  \nonumber  \\
 & & \,\,\, \exi{V} ((R\,T\,V) \wedge (R\,U\,V))) \label{24} 
\end{eqnarray}
Remember, that $R$ is of type $\worldtype \typearrow \worldtype \typearrow o$ and $S,T,U$ are of type $\worldtype$.
The corresponding axioms are given next. \\[-1em]
\begin{minipage}{.45\textwidth}
\begin{eqnarray}
M: & & \all{\phi} \mball{r}\phi \mimpl \phi \label{15a} \\
B: & & \all{\phi} \phi \mimpl \mball{r}\mdexi{r}\phi \label{16a} \\
D: & & \all{\phi} \mball{r}\phi \mimpl \mdexi{r}\phi \label{17a} \\
4: & & \all{\phi} \mball{r}\phi \mimpl \mball{r}\mball{r}\phi \label{18a} \\
5: & &  \all{\phi} \mdexi{r}\phi \mimpl \mball{r}\mdexi{r}\phi \label{19a} 
\end{eqnarray}
\end{minipage}
\hfill
\begin{minipage}{.52\textwidth}
\begin{eqnarray}
& &  \all{\phi} \mdexi{r}\phi \mimpl \mball{r}\phi \label{20a} \\
& &  \all{\phi} \mdexi{r}\phi \miff \mball{r}\phi \label{21a} \\
& &  \all{\phi} \mball{r}\mball{r}\phi \mimpl \mball{r}\phi \label{22a} \\
& &  \all{\phi,\psi} \mball{r}((\phi \mand \mball{r}\phi) \mimpl \psi) \mor \nonumber\\
& & \phantom{\all{\phi,\psi}}\mball{r}((\psi \mand \mball{r}\psi) \mimpl \phi) \label{23a} \\
& &  \all{\phi} \mdexi{r}\mball{r}\phi \mimpl \mball{r}\mdexi{r}\phi \label{24a}\\ \nonumber 
\end{eqnarray}
\end{minipage}

\begin{example}
For $k$ ($k = (\ref{15}), \ldots, (\ref{24})$) we can now easily formulate the well known correspondence theorems $(k) \Rightarrow (k+10)$ and $(k) \Leftarrow (k+10)$. For example,
\begin{eqnarray*}
(\ref{15}) \Rightarrow (\ref{15a}):\quad \all{R} (\text{reflexive}\;R) \Rightarrow (\text{valid}\; \all{\phi} \mball{R}\phi \mimpl \phi ) 
\end{eqnarray*}
\end{example}





\begin{example} There are well known relationships between different modal logics and there exist alternatives for their axiomatization (cf. the relationship map in \cite{sep-logic-modal}).
For example, for modal logic S5 we may choose axioms M and 5 as standard axioms. Respectively for logic KB5 we may choose B and 5. We may then want to investigate the following conjectures (the only one that does not hold is (\ref{D45impl5})):\\
\begin{minipage}{.5\textwidth}
\begin{eqnarray}
\text{S5}=\text{M5} & \Leftrightarrow & \text{MB5} \label{M5equivMB5} \\
   & \Leftrightarrow & \text{M4B5} \label{M5equivM4B5} \\
   & \Leftrightarrow & \text{M45} \label{M5equivM45} \\
   & \Leftrightarrow & \text{M4B} \label{M5equivM4B} \\
   & \Leftrightarrow & \text{D4B} \label{M5equivD4B} \\
   & \Leftrightarrow & \text{D4B5} \label{M5equivD4B5} \\
   & \Leftrightarrow & \text{DB5} \label{M5equivDB5} \\
\nonumber
\end{eqnarray}
\end{minipage}
\hfill
\begin{minipage}{.5\textwidth}
\begin{eqnarray}
\text{KB5} & \Leftrightarrow & \text{K4B5} \label{KB5equivK4B5} \\
    & \Leftrightarrow & \text{K4B} \label{KB5equivK4B} \\
\nonumber \\
 \text{M5} & \Rightarrow & \text{D45} \label{M5implD45} \\
 \text{D45} & \Rightarrow & \text{M5} \label{D45impl5} 
\end{eqnarray}
\end{minipage}
Exploiting the correlations $(k) \Leftrightarrow (k+10)$ from before these problems can be formulated as follows; we give the case for $\text{M5} \Leftrightarrow \text{D4B}$:
$$\all{R} (((\text{reflexive}\;R) \wedge (\text{euclidean}\;R)) \Leftrightarrow ((\text{serial}\;R) \wedge (\text{transitive}\;R) \wedge (\text{symmetric}\;R)))$$
\end{example}

\begin{example} We can also encode the Barcan formula and its converse. (They are theorems in our approach, which confirms that we are 'constant domain'.)
\begin{eqnarray}
BF:\quad  & \text{valid}\,\, \mall{X_\indtype} \mball{r} (p_{\indtype\typearrow(\worldtype\typearrow o)}\, X) \mimpl \mball{r} \mall{X_\indtype} (p_{\indtype\typearrow(\worldtype\typearrow o)}\, X) \label{Barcan1} \\
BF^{-1}:\quad & \text{valid}\,\, \mball{r} \mall{X_\indtype} (p_{\indtype\typearrow(\worldtype\typearrow o)}\, X) \mimpl \mall{X_\indtype} \mball{r} (p_{\indtype\typearrow(\worldtype\typearrow o)}\, X) \label{Barcan2}
\end{eqnarray}
\end{example}


\begin{example} An interesting meta property for combined logics with modalities $\modal\Diamond_{i},
\modal\Box_{j}, \modal\Box_{k},$ and $\modal\Diamond_{l}$ is the correspondence between the following  axiom  and the
$(i,j,k,l)$-confluence property
\begin{eqnarray}
&& (\text{valid}\,\, \mall{\phi} (\mdexi{i}
\mball{j} \phi) \mimpl \mball{k} \mdexi{l} \phi) \nonumber \\
& \Leftrightarrow &  (\all{A}
\all{B} \all{C} (((i\, A\, B) \wedge (k\, A\,
C)) \Rightarrow \exi{D} ((j\, B\, D) \wedge (l\, C\, D)))) \label{Correspondence2}
\end{eqnarray}
\end{example}

\begin{example}
Segerberg \cite{Segerberg73} discusses a 2-dimensional logic providing two S5 modalities
$\modal\Box_{a}$ and $\modal\Box_{b}$. He adds further axioms stating
that these modalities are commutative and orthogonal. It actually
turns out that orthogonality is already implied in this context. This statement can be encoded in our framework as follows:
\begin{eqnarray}
&& (\text{reflexive}\;a), (\text{transitive}\;a), (\text{euclid.}\;a), (\text{reflexive}\;b), (\text{transitive}\;b), (\text{euclid.
}\,b), \nonumber \\
&& (valid\; \mall{\phi} \mball{a} \mball{b} \phi  \miff \mball{b} \mball{a} \phi) \nonumber \\
&& \models^\STT\;(valid\; \mall{\phi,\psi} \mball{a}(\mball{a}\phi \vee \mball{b} \psi) \mimpl (\mball{a}\phi \vee \mball{a} \psi)) \wedge \nonumber \\
&& \phantom{\models^\STT\;\;}(valid\; \mall{\phi,\psi} \mball{b}(\mball{a}\phi \vee \mball{b} \psi) \mimpl (\mball{b}\phi \vee \mball{b} \psi))  \label{segerberg1}   
\end{eqnarray}
\end{example}

\begin{example}
Suppose we want to work with a 2-dimensional logic combining a
modality $\modal\Box_{k}$ of knowledge with a modality $\modal\Box_{b}$ of
belief. Moreover, suppose we model $\modal\Box_{k}$ as an S5 modality and $\modal\Box_{b}$
as an D45 modality and let us furthermore add two axioms characterizing
their relationship. We may then want to check whether or not
$\modal\Box_{k}$ and $\modal\Box_{b}$ coincide, i.e., whether $\modal\Box_{k}$  includes $\modal\Box_{b}$: 
\begin{eqnarray}
&& (\text{reflexive}\;k), (\text{transitive}\;k), (\text{euclid.}\;k), (\text{serial}\;b), (\text{transitive}\;b), (\text{euclid.}\,b), \nonumber \\
&& (valid\; \mall{\phi} \mball{k} \phi  \mimpl \mball{b} \phi), (valid\; \mall{\phi} \mball{b} \phi  \mimpl \mball{b} \mball{k} \phi) \nonumber \\
&& \models^\STT\; (valid\; \mall{\phi} \mball{b} \phi  \mimpl \mball{k} \phi) \label{knowledge1}   
\end{eqnarray}
\end{example}

\section{Reasoning within Combined Logics} \label{sec:within} 
We illustrate how our approach supports reasoning within combined logics.
First we present two examples in epistemic reasoning. Our
formulation in both cases adapts Baldoni's modeling \cite{baldoni03:_normal_multim_logic}.

\begin{example}[Epistemic reasoning:  The friends puzzle] \label{mmex} (i) Peter is
  a friend of John, so if Peter knows that John knows something then John knows
  that Peter knows the same thing. (ii) Peter is married, so if
  Peter's wife knows something, then Peter knows the same thing. John
  and Peter have an appointment, let us consider the following
  situation: (a) Peter knows the time of their appointment. (b) Peter
  also knows that John knows the place of their appointment. Moreover,
  (c) Peter's wife knows that if Peter knows the time of their
  appointment, then John knows that too (since John and Peter are
  friends). Finally, (d) Peter knows that if John knows the place and
  the time of their appointment, then John knows that he has an
  appointment.  From this situation we want to prove (e) that each of the two
  friends knows that the other one knows that he has an appointment. \\

For modeling the knowledge of Peter, Peter's wife, and John we consider a 3-dimensional logic combining
the modalities $\modal\Box_{\text{p}}{}$, $\modal\Box_{\text{(w\,p)}}{}$,
and $\modal\Box_{\text{j}}{}.$ Actually modeling them as S4 modalities turns out to be sufficient for this example. 
 Hence, we introduce three corresponding accessibility relations \text{j}, \text{p}, and
\text{(w\,p)}. The S4 axioms for $x\in\{\text{j}, \text{p},
\text{(w\,p)}\}$ are \\
\begin{minipage}{.4\textwidth}
\begin{eqnarray}
&& \text{valid}\; \mall{\phi} \mball{{x}} \phi \mimpl \phi \label{mm1} \\
\nonumber
\end{eqnarray}
\end{minipage}
\hfill
\begin{minipage}{.5\textwidth}
\begin{eqnarray}
&&  \text{valid}\; \mall{\phi} \mball{{x}} \phi \mimpl \mball{{x}} \mball{{x}} \phi \label{mm2} \\
\nonumber
\end{eqnarray}
\end{minipage}

\noindent As done before, we could alternatively postulate that the accessibility relations  are reflexive and transitive.

Next, we encode the facts from the puzzle. For (i) we provide a persistence axiom and for (ii) an inclusion axiom: \\
\begin{minipage}{.45\textwidth}
\begin{align}
&\text{valid}\;  \mall{\phi} \mball{\text{p}} \mball{\text{j}} \phi \mimpl \mball{\text{j}} \mball{\text{p}} \phi \label{mm3}\\
\nonumber
\end{align}
\end{minipage}
\hfill
\begin{minipage}{.45\textwidth}
\begin{align}
&\text{valid}\;  \mall{\phi} \mball{\text{(w\,p)}} \phi \mimpl \mball{\text{p}} \phi \label{mm4} \\
\nonumber
\end{align}
\end{minipage}

Finally, the facts (a)-(d) and the conclusion (e) are encoded as follows (time, place, and appointment are propositional constants, that is, constants of type $\worldtype\typearrow o$ in our framework):
\begin{align}
 &\text{valid}\;  \mball{\text{p}} \text{time} \label{mm5}\\
 &\text{valid}\;  \mball{\text{p}} \mball{\text{j}} \text{place} \label{mm6}\\ 
 &\text{valid}\;  \mball{\text{(w\,p)}} (\mball{\text{p}} \text{time} \mimpl \mball{\text{j}} \text{time}) \label{mm7}\\ 
 &\text{valid}\;  \mball{\text{p}} \mball{\text{j}} (\text{place} \mand \text{time} \mimpl \text{appointment}) \label{mm8} \\
 &\text{valid}\;  \mball{\text{j}} \mball{\text{p}} \text{appointment} \mand \mball{\text{p}} \mball{\text{j}} \text{appointment} \label{mm9}
\end{align}
The combined proof problem for Example \ref{wise} is 
\begin{eqnarray}
(\ref{mm1}),\ldots,(\ref{mm8}) \models^\stt (\ref{mm9}) \label{example1}
\end{eqnarray}
\end{example}

\begin{example}[Wise men puzzle]\label{wise}
Once upon a time, a king wanted to find the wisest out of his three wisest
men. He arranged them in a circle and told them that he would put a white or a black spot on
their foreheads and that one of the three spots would certainly be white. 
 The three wise men could see and hear each other but, of course, they could not see their faces reflected anywhere.
 The king, then, asked to each of them to find out the color of his own spot. After a while, the
wisest correctly answered that his spot was white.\\

We employ a 4-dimensional logic combining the modalities  $\modal\Box_{\text{a}}$, $\modal\Box_{\text{b}}$,
and $\modal\Box_{\text{c}}$, for encoding the individual knowledge of the
three wise men, and a box operator $\modal\Box_{\text{fool}}$, for encoding
the knowledge that is common to all of them.  The entire encoding consists now of the
following axioms for $X,Y,Z \in \{a,b,c\}$ and $X\not=Y\not=Z$:

\begin{eqnarray}
&& \text{valid}\; \mball{\text{fool}} ((\text{ws}\;\text{a}) \mor (\text{ws}\;\text{b}) \mor (\text{ws}\;\text{c})) \label{ax1}\\
&& \text{valid}\; \mball{\text{fool}} ((\text{ws}\;X) \mimpl \mball{Y} (\text{ws}\;X)) \label{ax2}\\
&& \text{valid}\; \mball{\text{fool}} (\mnot (\text{ws}\;X) \mimpl \mball{Y} \mnot (\text{ws}\;X)) \label{ax3}\\
&& \text{valid}\; \mall{\phi} \mball{\text{fool}} \phi \mimpl \phi \label{ax4}\\
&& \text{valid}\; \mall{\phi} \mball{\text{fool}} \phi \mimpl \mball{\text{fool}}\mball{\text{fool}}\phi \label{ax5}\\
&& \text{valid}\; \mall{\phi} \mball{\text{fool}} \phi \mimpl \mball{\text{a}}\phi \label{ax6}\\
&& \text{valid}\; \mall{\phi} \mball{\text{fool}} \phi \mimpl \mball{\text{b}}\phi \label{ax7}\\
&& \text{valid}\; \mall{\phi} \mball{\text{fool}} \phi \mimpl \mball{\text{c}}\phi \label{ax8}\\
&& \text{valid}\; \mall{\phi} \mnot \mball{\text{X}} \phi \mimpl \mball{\text{Y}} \mnot \mball{\text{X}} \phi \label{ax9}\\
&& \text{valid}\; \mall{\phi} \mball{\text{X}} \phi \mimpl \mball{\text{Y}} \mball{\text{X}} \phi \label{ax9b}\\
&& \text{valid}\; \mnot \mball{\text{a}} (\text{ws}\;\text{a}) \label{ax10}\\
&& \text{valid}\; \mnot \mball{\text{b}} (\text{ws}\;\text{b}) \label{ax11}
\end{eqnarray}
From these assumptions we want to conclude that 
\begin{eqnarray}
&& \text{valid}\; \mball{\text{c}} (\text{ws}\;\text{c})  \label{ax12}
\end{eqnarray}

Axiom (\ref{ax1}) says that a, b, or c must have a white spot and that
this information is known to everybody. Axioms (\ref{ax2}) and
(\ref{ax3}) express that it is generally known that if someone has a
white spot (or not) then the others know this.  $\modal\Box_{\text{fool}}$
is axiomatized as an S4 modality in axioms (\ref{ax4}) and
(\ref{ax5}). For $\modal\Box_{\text{a}}$, $\modal\Box_{\text{b}}$, and
$\modal\Box_{\text{c}}$ it is sufficient to consider K modalities. The
relation between those and common knowledge ($\modal\Box_{\text{fool}}$ modality) is
axiomatized in inclusion axioms (\ref{ax6})--(\ref{ax9}).  Axioms
(\ref{ax9}) and (\ref{ax9b}) encode that whenever a wise man does
(not) know something the others know that he does not know
this. Axioms (\ref{ax10}) and (\ref{ax11}) say that a and b do not
know whether they have a white spot. Finally, conjecture (\ref{ax12})
states that that c knows he has a white spot. The combined proof
problem for Example \ref{mmex} is
\begin{eqnarray}
(\ref{ax1}),\ldots,(\ref{ax11}) \models^\stt (\ref{ax12}) \label{example2}
\end{eqnarray}
\end{example}

\begin{example}
A trivial example problem for the region connection calculus is (adapted from \cite{GabbayKuruczWolterZakharyaschev2002}, p. 80):
\begin{eqnarray}
& & (tpp\;\text{catalunya}\;\text{spain}), \label{RCC1ex2.ax} \nonumber \\
& & (ec\;\text{spain}\;\text{france}), \label{RCC1ex3.ax} \nonumber \\
& & (ntpp\;\text{paris}\;\text{france}), \label{RCC1ex4.ax} \nonumber \\
& & \models^\stt  (dc\;\text{catalunya}\;\text{paris}) \wedge (dc\;\text{spain}\;\text{paris}) \label{RCCex1.p}
\end{eqnarray}
\end{example}
The assumptions express that (i) Catalunya is a border region of Spain, (ii) Spain and France are two different countries sharing a common border, and (iii) Paris is a proper part of France. The conjecture is that (iv) Catalunya and Paris are disconnected as well as Spain and Paris.

\begin{example}
Within our $\stt$ framework we can easily put such spatial reasoning examples in an epistemic context; similar to before we distinguish between common knowledge ($\text{fool}$) and the knowledge of person $\text{bob}$ and we lift the above propositions to modal propositions of type $\worldtype \typearrow o$:
\begin{eqnarray}
&& \text{valid}\; \mall{\phi} \mball{\text{fool}} \phi \mimpl \mball{\text{bob}}\phi,\nonumber\\
& & \text{valid}\; \mball{\text{bob}} (\lam{W} (tpp\;\text{catalunya}\;\text{spain})),\nonumber \\
& & \text{valid}\; \mball{\text{fool}} (\lam{W} (ec\;\text{spain}\;\text{france})),\nonumber \\
& & \text{valid}\; \mball{\text{bob}} (\lam{W} (ntpp\;\text{paris}\;\text{france})) \nonumber \\
& & \models^\stt \nonumber \\
& & \text{valid}\; \mball{\text{bob}} (\lam{W} ((dc\;\text{catalunya}\;\text{paris}) \wedge (dc\;\text{spain}\;\text{paris})))  \label{EpistemicRCCex1.p} 
\end{eqnarray}
We here express that (ii) from above is commonly known, while (i) and (ii) are not. (i) and (ii) are known to the educated person bob though. In this situation, conjecture (iv) still follows for bob. However, it does not follow when replacing bob by common knowledge (hence, the following problem is not provable):
\begin{eqnarray}
& & \ldots \models^\stt  \label{EpistemicRCCex2.p}  \text{valid}\; \mball{\text{fool}} (\lam{W}  ((dc\;\text{catalunya}\;\text{paris}) \wedge (dc\;\text{spain}\;\text{paris})))
\end{eqnarray}
\end{example}

\section{Experiments} \label{sec:exp} In our case studies, we have
employed the $\stt$ automated reasoners LEO-II---v1.1 \cite{BP+08},
TPS---3.080227G1d \cite{DBLP:journals/japll/AndrewsB06},
IsabelleP---2009-1, IsabelleM---2009-1, and
IsabelleN---2009-1\@.\footnote{IsabelleM and IsabelleN are model finder
  in the Isabelle proof assistant \cite{DBLP:books/sp/NipkowPW02} that
  have been made available in batch mode, while IsabelleP applies a
  series of Isabelle proof tactics in batch mode.}  These systems are
available online via the SystemOnTPTP tool \cite{Sut07-CSR} and they
support the new TPTP THF infrastructure for typed higher-order logic
\cite{BRS08}.

The axiomatizations of $\QMLSTT$ and $\IPLSTT$ are available as
{LCL013\^{}0.ax} and {LCL010\^{}0.ax} in the TPTP
library.\footnote{Note that the types $\worldtype$ and $\indtype$ are unfortunately switched in the encodings available in the TPTP: the former is used for individuals and the latter for worlds. This syntactic switch is completely unproblematic.}  The example problems {LCL698\^{}1.p} and {LCL695\^{}1.p}
ask about the satisfiability of these axiomatizations.  Both questions
are answered positively by IsabelleM and IsabelleN; IsabelleM needs 3.8
resp. 3.6 seconds and IsabelleN 3.8 resp. 3.6 seconds.

Table \ref{table1} presents the results of our experiments; the
timeout was set to 120 seconds and the entries in the table are
reported in seconds. 
Those examples which have already entered the new higher-order TPTP library
are presented with their respective TPTP identifiers in the second column and the others will soon be submitted.
\begin{table}[htp]
\begin{center}
\begin{tabularx}{0.8\textwidth}{|X|X|X|X|X|} 
\hline
Problem        & TPTP id             & LEO-II  & TPS  & IsabelleP \\
\hline 
\hline
\multicolumn{5}{|c|}{Reasoning about Logics and Combined Logics} \\
(\ref{15}) $\Rightarrow$ (\ref{15a}) & LCL699\^{}1.p & 0.0 & 0.3 & 3.6 \\ 
(\ref{16}) $\Rightarrow$ (\ref{16a}) & LCL700\^{}1.p & 0.0 & 0.3 & 13.9 \\ 
(\ref{17}) $\Rightarrow$ (\ref{17a}) & LCL701\^{}1.p & 0.0 & 0.3 & 4.0 \\ 
(\ref{18}) $\Rightarrow$ (\ref{18a}) & LCL702\^{}1.p & 0.0 & 0.3 & 15.9 \\  
(\ref{19}) $\Rightarrow$ (\ref{19a}) & LCL703\^{}1.p & 0.1 & 0.3 & 16.0 \\
(\ref{20}) $\Rightarrow$ (\ref{20a}) & LCL704\^{}1.p & 0.0 & 0.3 & 3.6 \\  
(\ref{21}) $\Rightarrow$ (\ref{21a}) & LCL705\^{}1.p & 0.1 & 51.2 & 3.9 \\  
(\ref{22}) $\Rightarrow$ (\ref{22a}) & LCL706\^{}1.p & 0.1 & 0.3 & 3.9 \\  
(\ref{23}) $\Rightarrow$ (\ref{23a}) & LCL707\^{}1.p & 0.1 & 0.3 & 3.6 \\
(\ref{24}) $\Rightarrow$ (\ref{24a}) & LCL708\^{}1.p & 0.1 & 0.3 & 4.1 \\

(\ref{15}) $\Leftarrow$ (\ref{15a})  & LCL709\^{}1.p & 0.0 & 0.3 & 3.7 \\
(\ref{16}) $\Leftarrow$ (\ref{16a})  & LCL710\^{}1.p & --- & 0.3 & 53.8  \\
(\ref{17}) $\Leftarrow$ (\ref{17a})  & LCL711\^{}1.p & 0.0 & 0.3 & 3.7 \\ 
(\ref{18}) $\Leftarrow$ (\ref{18a})  & LCL712\^{}1.p & 0.0 & 0.3 & 3.8 \\ 
(\ref{19}) $\Leftarrow$ (\ref{19a})  & LCL713\^{}1.p & --- & 0.8 & 67.0 \\ 
(\ref{20}) $\Leftarrow$ (\ref{20a})  & LCL714\^{}1.p & 1.6 & 0.3 & 29.3 \\ 
(\ref{21}) $\Leftarrow$ (\ref{21a})  & LCL715\^{}1.p & 37.9 & --- & --- \\ 
(\ref{22}) $\Leftarrow$ (\ref{22a})  & LCL716\^{}1.p & --- & 6.6 & --- \\ 
(\ref{23}) $\Leftarrow$ (\ref{23a})  & LCL717\^{}1.p & --- & --- & --- \\ 
(\ref{24}) $\Leftarrow$ (\ref{24a})  & LCL718\^{}1.p & 0.1 & 0.4 & 8.1 \\
 
(\ref{M5equivMB5}) & & 0.1 & 0.4 & 4.3 \\
(\ref{M5equivM4B5}) & & 0.2 & 27.4 & 4.0 \\
(\ref{M5equivM45}) & & 0.1 & 8.9 & 4.0 \\
(\ref{M5equivM4B}) & & 0.1 & 1.2 & 3.7 \\
(\ref{M5equivD4B}) & & 0.1 & 1.7 & 4.2 \\
(\ref{M5equivD4B5}) & & 0.2 & 14.8 & 5.4 \\
(\ref{M5equivDB5}) & & 0.1 & 0.6 & 3.7 \\
(\ref{KB5equivK4B5}) &  & 0.2 & 2.3 & 4.0 \\
(\ref{KB5equivK4B}) & & 0.1 & 0.9 & 3.9 \\
(\ref{M5implD45}) & & 0.1 & 12.8 & 16.5 \\

(\ref{D45impl5})$^\text{Countersatisfiable}$ & & --- & --- & --- \\

(\ref{Barcan1})   &  & 0.0 & 0.3 & 3.6 \\
(\ref{Barcan2})   &  & 0.0 & 0.3 & 3.6 \\
(\ref{Correspondence2}) &  & 0.1 & 0.4 & 3.6 \\
(\ref{segerberg1}) &  & 0.2 & 35.5 & --- \\  
(\ref{knowledge1}) &  & 0.4 & --- & ---  \\  
\hline
\multicolumn{5}{|c|}{Reasoning within Combined Logics} \\
(\ref{example1}) & PUZ086\^{}1.p & 0.1 & --- & 102.4 \\ 
(\ref{example2}) & PUZ087\^{}1.p & 0.3 & --- & --- \\ 
(\ref{RCCex1.p})   & & 2.3 & --- & 112.7  \\
(\ref{EpistemicRCCex1.p}) & & 20.4 & --- & --- \\
(\ref{EpistemicRCCex2.p})$^\text{Countersatisfiable}$ & & ---   & ---   & --- \\
\hline
\end{tabularx}
\end{center}
\caption{Performance results of $\stt$ provers for problems in paper. \label{table1}}
\end{table}

As expected, (\ref{D45impl5}) and (\ref{EpistemicRCCex2.p}) cannot be
proved by any prover and IsabelleN reports a
counterexample for (\ref{D45impl5}) in 34.4 seconds and for
(\ref{EpistemicRCCex2.p}) in 39.7 seconds.

In summary, all but one of our example problems can be solved
effectively by at least one of the reasoners. In fact, most of
our example problems require only milliseconds.  LEO-II solves most problems
and it is the fastest prover.

\section{Conclusion} 
\label{Conclusion}
The work presented in this paper has its roots in the LEO-II project
(in 2006/2007 at University of Cambridge, UK) in which we first studied and
employed the presented embedding of quantified multimodal logics in
$\stt$ \cite{BP09}. 

Our overall goal is to show that various interesting classical and
non-classical logics and their combinations can be elegantly mechanized and
partly automated in modern higher-order reasoning systems with
the help of our logic embeddings.

Our experiments are encouraging and they provide first evidence for
our claim that $\stt$ is suited as a framework for combining
classical and non-classical logics.  It is obvious, however, that
$\stt$ reasoners should be significantly improved for fruitful
application to more challenge problems in practice.  The author is
convinced that significant improvements --- in particular for
fragments of $\stt$ as illustrated in this paper --- are possible and
that they will be fostered by the new TPTP infrastructure and the new
yearly higher-order CASC competitions.

Moreover, when working with our reasoners from within a proof
assistant such as Isabelle/HOL the user may provide interactive help,
for example, by formulating some lemmas or by splitting proof tasks in
simpler subtasks.  

An advantage of our approach also is that provers such as our LEO-II
are generally capable of producing verifiable proof output, though
much further work is needed to make these proof protocols exchangeable
between systems or to explain them to humans.  Finally note that it may
be possible to formally verify the entire theory of our embedding(s)
within a proof assistant.

\paragraph{Acknowledgment:} 
The author is indebted to Geoff Sutcliffe, who, in collaboration with
the author and supported by several further contributors, developed the new
higher-order TPTP THF infrastructure in the EU FP7 Project THFTPTP (grant
PIIF-GA-2008-219982).




\bibliographystyle{plain} 

\end{document}